\documentclass[pra,twocolumn,showpacs,preprintnumbers,amsmath,amssymb]{revtex4}
\usepackage{graphicx}% Include figure files
\usepackage{dcolumn}% Align table columns on decimal point
\usepackage{bm}% bold math
\begin{document}
\title{An Investigation of Mean-field Effects for a {B}ose Condensate in an Optical Lattice}

\author{S.~B.~McKagan$^1$$^,$\footnote{Current address: JILA, University of Colorado, Boulder, CO 80309}, D.~L.~Feder$^2$, and W.~P.~Reinhardt$^1$}
\affiliation{$^1$Departments of Chemistry and Physics, University of
Washington, Seattle, WA 98195}
%\affiliation{$^2$JILA, University of Colorado, Boulder, CO 80309 (current address)}
\affiliation{$^2$Department of Physics and Astronomy and the Institute for
Quantum Information Science, University of Calgary, Calgary, Alberta, Canada
T2N 1N4}

\date{\today}

\begin{abstract}
This paper presents a mean-field numerical analysis, using the full
three-dimensional time-dependent Gross-Pitaevskii equation (GPE), of an
experiment carried out by Orzel {\it et al.} [{\it Science}~\textbf{291},
2386 (2001)] intended to show number squeezing in a gaseous Bose-Einstein
condensate in an optical lattice. The motivation for the present work is to
elucidate the role of mean-field effects in understanding the experimental
results of this work and those of related experiments.
We show that the non-adiabatic loading of atoms into optical lattices
reproduces many of the main results of the Orzel {\it et al.} experiment,
including both loss of interference patterns as laser intensity is increased
and their regeneration when intensities are lowered. The non-adiabaticity
found in the GPE simulations manifests itself primarily in a coupling
between the transverse and longitudinal dynamics, indicating that
one-dimensional approximations are inadequate to model the experiment.
\end{abstract}

\pacs{03.75.Dg, 03.75.Lm, 05.30.Jp, 32.80.Pj}% PACS, the Physics and Astronomy
\maketitle

\section{Introduction}
The creation of dilute gaseous Bose-Einstein condensates (BECs) in the
laboratory in 1995 \cite{Anderson1995a,Bradley1995a,Davis1995b} has spurred
much development in both experiment and theory
\cite{Dalfovo1999a,Leggett2001a,Pethick2002a}.  Mean-field theory was able
to explain most early BEC experiments, using the well-known Gross-Pitaevskii
equation (GPE)~\cite{Dalfovo1999a,Leggett2001a} which is valid for weakly
interacting BECs at zero temperature because it assumes all atoms occupy a
single macroscopic wavefunction:
\begin{equation}
i\hbar\frac{\partial}{\partial t}\psi(\mathbf{r},t)
=\left(-\frac{\hbar^2}{2m}\nabla^2 + V(\mathbf{r},t)
+g|\psi(\mathbf{r},t)|^2\right)\psi(\mathbf{r},t)
\label{gpe}
\end{equation}
where $\psi(\mathbf{r},t)$ is the single particle wave function for any atom
in the BEC, $V(\mathbf{r},t)$ is an external trapping potential, and
$g=4\pi a_s\hbar^2/m$, where $a_s$ is the s-wave scattering length and $m$ is
the atomic mass. These early successes included the anisotropic profile and
momentum distribution of the ground state after ballistic
expansion~\cite{Anderson1995a,Davis1995b,Mewes1996}, the spectrum of
collective excitations~\cite{Jin1996a,StamperKurn1998}, the dynamics of spinor
condensates~\cite{Miesner1999,Matthews1999a},
and the wave interference of interacting condensates
\cite{Andrews1997b,Rohrl1997a}.  Perhaps unexpectedly, the time dependent GPE
also accurately described strongly nonlinear excitations such as
vortices~\cite{Matthews1999b,Madison2000,AboShaeer2001} and
solitons~\cite{Burger1999a,Denschlag2000a}.

Given the overwhelming success of the GPE, there has been much interest in
finding situations in which it breaks down and a more detailed theoretical
description is needed. Indeed, the theoretical interpretation of the earliest
experiments on collective excitations at finite
temperature~\cite{Jin1997,StamperKurn1998,Marago2001} has required a dynamical
theory that includes the motion of the noncondensate~\cite{Jackson2002b,Morgan2003}.
More recent experiments are geared toward inducing strong correlations
among atoms, which are not captured by the GPE; these include molecules in
Bose~\cite{Wynar2000,McKenzie2002} and
Fermi~\cite{Greiner2003,Jochim2003,Zweierlein2003} gases and exploring the
BCS-BEC crossover~\cite{Regal2004,Bartenstein2004,Zweierlein2004},
low-dimensional systems through tight
confinement~\cite{Moritz2003,Tolra2004,Stoferle2004}, quantum Hall-like states
in rapidly rotating traps~\cite{Wilkin2000,Paredes2001} or using external
lasers~\cite{Jaksch2003}, and novel many-body states of atoms in optical
lattices~\cite{Anderson1998a,Orzel2001a,Greiner2002a,Stoferle2004,Paredes2004,Fertig2005,Xu2005}.

In the experiments with optical lattices, the extent of correlations in the
atomic gas has generally been measured by dropping all confining potentials
and imaging the cloud after a period of ballistic expansion. When condensates
are suddenly released from their confinement in shallow optical lattices
formed from weak lasers, they form well-defined interference patterns
corresponding to momentum-space Bragg peaks. This has been interpreted as an
unequivocal signature of phase coherence over multiple lattice sites. As the
laser intensity increases, however, the interference patterns partially or
fully wash out~\cite{Anderson1998a,Orzel2001a,Greiner2002a}. The theoretical
interpretation is that as tunneling is quenched, the initial macroscopic
condensate splits into many separate sub-condensates which may then lose some
or all of their relative coherence; the resulting state has been described as
`number squeezed,' or `fragmented.' Under various conditions one obtains
full fragmentation, which leads to a quantum phase transition from a
superfluid state to a Mott insulator~\cite{Fisher1989}.

In fact, the loss of fringe contrast after ballistic expansion is not as clear
a signature of condensate fragmentation as is often assumed. Jean Dalibard and
collaborators~\cite{Hadzibabic2004} recently demonstrated that
high-visibility interference patterns resulted even under conditions where
the phases from site to site of a one-dimensional (1D) optical lattice were
random. A simple theoretical model that generalizes the interference pattern
from two uncorrelated sources reproduces the experimental data. Under
conditions similar to those of Orzel {\it et al.}~\cite{Orzel2001a} where
number squeezing is expected to be present (though using a blue-detuned rather
than a red-detuned lattice where the transverse confinement is much weaker),
they observed an unattributed heating effect. A similar heating was observed
by Morsch {\it et al.}~\cite{Morsch2003}, where radial modes excited by the
non-adiabatic ramp-up of the 1D optical lattice rapidly damped and transfered
energy to high-lying axial modes.  Zakrzewski~\cite{Zakrzewski2005a}, using a time-dependent Gutzwiller mean-field approach to solving the Bose-Hubbard model, has found that non-adiabatic mean field effects reproduce some of the results seen in the experiments of Greiner {\it et al.}~\cite{Greiner2002a}.

The recent experiments indicate that fundamental questions need to be answered
before fringe contrast can be used as a
quantitative measure of the breakdown of mean-field theory in these systems.
The effects of non-adiabaticity on the phase variations across the lattice have
been addressed previously~\cite{Sklarz2002,Band2002b,Plata2004,Isella2005}. 1D
calculations based on the GPE found that the harmonically trapped condensate
develops a pronounced `phase sag' after slowly ramping up the
lattice~\cite{Sklarz2002,Band2002b}; the phase was found to increase
approximately quadratically around the trap center. A flat phase profile at
the end of the ramp, and a high-contrast interference pattern, can be restored
if the harmonic trap parameters are simultaneously varied~\cite{Band2002b}.
More recent effective 3D calculations demonstrate that collective modes are
excited by slow ramps~\cite{Band2003,Plata2004}, but the influence of these
excitations on the resulting interference patterns was not addressed.

In order to clarify the various mean-field effects that can degrade the
interference pattern when a condensate is released from an optical lattice,
we have performed fully 3D numerical simulations based on the GPE of one of
the above experiments, namely that of Orzel, Tuchman, Fenselau, Yasuda, and
Kasevich~\cite{Orzel2001a} (referred to as OTFYK in what follows). See
also~\cite{McKinney2004a}.  If number squeezing were the main reason for the
observed loss of fringe contrast as has been suggested~\cite{Orzel2001a}, the
GPE simulations
should not be able to mimic the experimental data. We find, however, that we
can reproduce the degradation of the interference patterns as the lattice depth
approaches the regime where squeezing would be expected, and the subsequent
restoration of contrast as the laser intensity is lowered, even in cases
where `random' phases have been applied to the individual wells at the largest
lattice depths.  At the level of the mean-field approximation, the behavior
is due to the non-adiabatic response of the trapped condensate as
the optical lattice is slowly turned on: the ramp induces axial currents and
the excitation of strongly coupled transverse and longitudinal excitations.
These in turn yield variations in the phase from site to site that cause the
interference patterns to disappear. We expect that in the presence of damping
(not included in the present analysis), the mean-field results would be
close to, if not indistinguishable from, the experimental data.

The paper is organized as follows:  In Section II, the ground states are
obtained for a condensate in harmonic oscillator, gravitational, and optical
lattice potentials as a function of laser intensity. The OTFYK experiment is
described in Section III, and the observations are compared to results of a
GPE model that assumes perfect adiabaticity of the optical lattice loading.
Section IV, which is the heart of the paper, contains discussion of a GPE
simulation of the full experimental protocol, including the turning on and
turning off of the lattice, and the use of gravity to induce a relative phase
shift of $\pi$ between adjacent lattice sites. It is found that the
experimental timescales for the lattice ramp are not generally sufficient
to ensure adiabaticity. The resulting phase shifts mimic loss of coherence
between adjacent lattice sites. In Sections V and VI it is demonstrated that
interference patterns recover as the laser intensities are reduced, even if
the site-to-site phases are artificially randomized at lattice maximum.
In addition to the accumulation of axial phase variations, a second source of
non-adiabaticity is found and discussed in Sections VII and VIII: a strong
coupling between longitudinal and transverse oscillations. The results are
compared to more recent experiments by the Kasevich
group~\cite{Tuchman2004a,Tuchman2005}.
A brief summary and conclusions end the paper, in Section IX.

\section{The Potential and Ground State Densities}
In the OTFYK experiment, a BEC is first created in a cylindrically symmetric
magnetic trap with the longitudinal axis oriented vertically, parallel to the
ambient gravitational field.  Counter-propagating laser beams are then slowly
turned on, oriented vertically and aligned with the long axis of the magnetic
trap.  The lasers are red detuned, and thus the condensate is pulled into the
anti-nodes, with spacing $\lambda/2$, and also experiences strong transverse
confinement.  The full potential used to describe in the present numerical
studies of this system is given by~\cite{Chiofalo2000a}:
\begin{eqnarray}\label{3dpot}
V(\rho,z,t)=U(t)\left(1-e^{-\rho^2/r_{lb}^2}\cos^2(kz)\right)\nonumber\\
+ \frac{1}{2}m({\omega}_{\perp}^2\rho^2+{\omega}_z^2z^2)+mgz
\end{eqnarray}
where the first term is due to the laser, the second term to the magnetic trap,
and the third term to gravity.  The longitudinal coordinate $z$ is in the
vertical direction and the transverse coordinate $\rho=\sqrt{x^2+y^2}$.
Table~\ref{parameters} gives the values of the experimental parameters as
reported in the OTFYK paper.

\begin{table}[t]
 \centering
 \caption{Values of parameters used in the experiment and
modeling.}\label{parameters}
 \small
 \begin{tabular}{|c|c|c|}
   \hline
   $N$ & 30000 & number of atoms in BEC\\
   $a_s$  & $5.24$nm~\cite{vanKempen2002a} & scattering length of $^{87}$Rb \\
   $m$  & $1.44\times10^{-25}$kg~\cite{Bradley1999a} & atomic mass of $^{87}$Rb
\\
   ${\omega}_{\perp}$ & $2\pi\times19$Hz & transverse frequency of trap \\
   ${\omega}_z$ & $2\sqrt{2}{\omega}_{\perp}$ & longitudinal frequency of trap
     \\
   $k$ & $2\pi/840$nm & wave number of laser \\
   $r_{lb}$ & $50\mu$m & 1/e intensity radius of laser \\
   $U$ & $0$ to $44E_R$ & Laser intensity ($E_R=\hbar^2k^2/2m$) \\
   \hline
  \end{tabular}
\end{table}

Various parts of the trap are turned on and off at various points in the
experiment, as described in Section III, so not all the terms in
Eq.~(\ref{3dpot}) are present at all times.  The laser is generally ramped up
slowly, so $U$ varies as a function of time.  As long as the magnetic trap
is on, the gravitational term has no effect other than to give a linear shift
the quadratic
trapping potential, but it plays an important role when the magnetic trap is
turned off.  In our calculations, we neglect this term except when it is
required for phase manipulation. Note that the laser creates not only a
periodic trapping potential in the longitudinal direction, but also a Gaussian
trapping potential in the transverse direction.  This Gaussian term (which is
not present for a blue-detuned laser) implies that as the laser strength
increases, so does the transverse confinement of the BEC, and therefore the
density increases.  This turns out to be a very important effect.

Although the
actual OTFYK experiments, as seen in Section III, consist of a sequence of well
defined time dependent steps, it is useful to examine the density profiles of
the stationary GPE ground states in the potential of Eq.~(\ref{3dpot}). Ground
states were obtained using numerical methods described in the Appendix, by
evolution of the GPE in complex time ($t\rightarrow it$) with propagation
converging to the lowest energy stationary states for each fixed value of
$U$. The results for three representative values of $U$ are shown in
Fig.~\ref{groundstates}. Figure~\ref{groundstates}(a) illustrates the density
profile in the absence of the optical lattice. All experiments and theoretical
simulations described herein start with this ground state of the condensate in
the magnetic trap in the presence of gravity. Figures~\ref{groundstates}(b) and
(c) illustrate the two dominant effects of the laser fields: first the
condensate morphs from a single
ellipsoidal and coherent structure into a vertical stack of disk like sub-
condensates. Also clearly seen is the very large effect of the transverse
confinement: the condensate density envelope changes from oblate to prolate
at the highest field intensities.

\section{An Initial ``Adiabatic" Simulation of the Experiment}
\begin{figure}[t]
\centering
\includegraphics[width=\columnwidth]{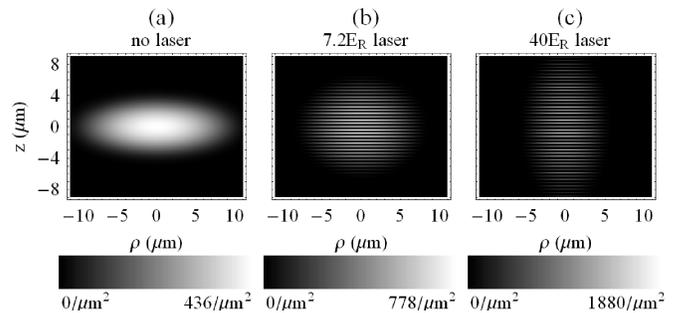}
\caption{Cross sections of integrated density for the ground state of a BEC in
a harmonic trap and laser, from complex time evolution.  Note that the effect
of the transverse confinement of the laser is very large, turning an initially
disk-shaped BEC into a cigar-shaped BEC.}\label{groundstates}
\end{figure}

We have simulated each step of the OTFYK experiment using the full 3D GPE with
the potential discussed in the previous section. To avoid repetition, we will
describe the steps of the experiment and our simulation in parallel.  The first
step of the experiment is to create a BEC in a magnetic trap.  This step can be
simulated by using complex time evolution as described in the previous section.
The results of the simulation for the harmonic trap only are shown in
Fig.~\ref{groundstates}(a).

The next step in the experiment involved turning on the laser fields so that
the intensity increases linearly from zero to some final intensity $U_f$
(ranging from $7.2E_R$ to $44E_R$) in a ramp time $\tau_R=200$~ms. This ramp-up
time was assumed by OTFYK  to be slow enough to allow the condensate to follow
the ramping adiabatically, so that the BEC stays in the appropriate ground
state as the laser is turned up~\cite{comment}; in the next section we will
test this assumption of adiabaticity. If the final state of the BEC
in the combined magnetic trap and lattice is the ground state of this system,
then the sequence of states generated are just those which may be found by
complex time evolution, as discussed in Section II. Then the ground states as
a function of $U$ shown in Figs.~\ref{groundstates}(b,c) can be used in a
preliminary simulation of the OTFYK experiment.

\begin{figure}[t]
\centering
\includegraphics[width=\columnwidth]{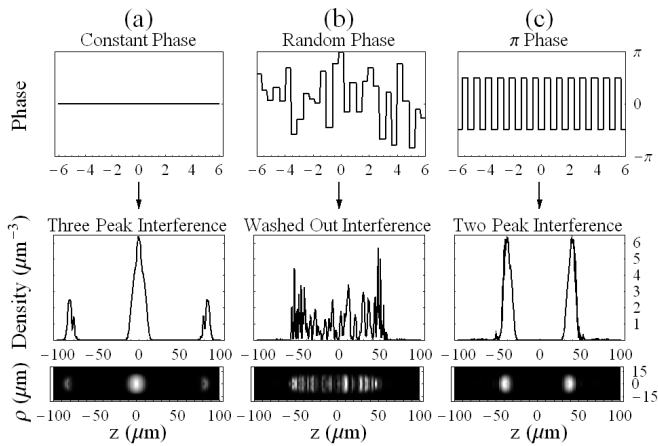}
\caption{Numerical simulations of three peak, washed out, and two
peak interference patterns produced by imprinting phases on ground state of GP
wave function in harmonic trap and $7.2E_R$ laser.  The first row shows the
phase along the longitudinal axis before expansion.  The second row shows the
density along the longitudinal axis after releasing the BEC and allowing it to
expand for $8$~ms.  The third row shows a cross section of the density after
expansion (indicated by the gray scale).  Note the difference in the scale
before and after expansions.}\label{expansions}
\end{figure}

The OTFYK paper described two regimes that exhibited markedly different
interference patterns after the external potentials were dropped and the
atomic cloud was allowed to ballistically expand. For low $U_f$, the final
state was expected to be fully phase coherent yielding a clean interference
pattern~\cite{Pedri2001}. The computed three-peak pattern for the $U_f=7.2E_R$
case is shown
in Fig.~\ref{expansions}(a). For high $U_f\sim 40E_R$, the condensate was
anticipated to be strongly number squeezed, leading to a random relative
phase from site to site. Ballistic expansion would then yield no detectable
interference pattern. Of course, the GPE cannot produce a number squeezed
state, but if the random phases are put in `by hand' it can mimic the
experimental situation, as was done previously for the MIT interference
experiment~\cite{Andrews1997b} by R\" ohrl {\it et al.}~\cite{Rohrl1997a}
within the framework of the GPE. The numerical results for this case are shown
in Fig.~\ref{expansions}(b).

The three peak interference pattern of Fig.~\ref{expansions}(a) is inconvenient for quantitative estimation of the extent of decay of coherence, so a third protocol was invoked by OTFYK. The experimenters used the presence of gravity, via an appropriate time delay, as discussed in the following section, to imprint a $\pi$ phase difference between the condensate in each pair of neighboring wells.  This produces a two-peak interference pattern, as illustrated in the simulated ballistic interference pattern of Fig.~\ref{expansions}(c).

The three
situations of Fig.~\ref{expansions} were simulated by painting phase patterns
`by hand' onto the ground state GP wave function in a $7.2E_R$ lattice and
then allowing it to spatially expand by turning off all external potentials.
The next section will show the condensate phase patterns obtained by a full
time-dependent simulation of the actual experiment.

\section{Full Time-dependent GPE Simulation of the OTFYK Experiment}
\begin{figure}[t]
\centering
\includegraphics[width=\columnwidth]{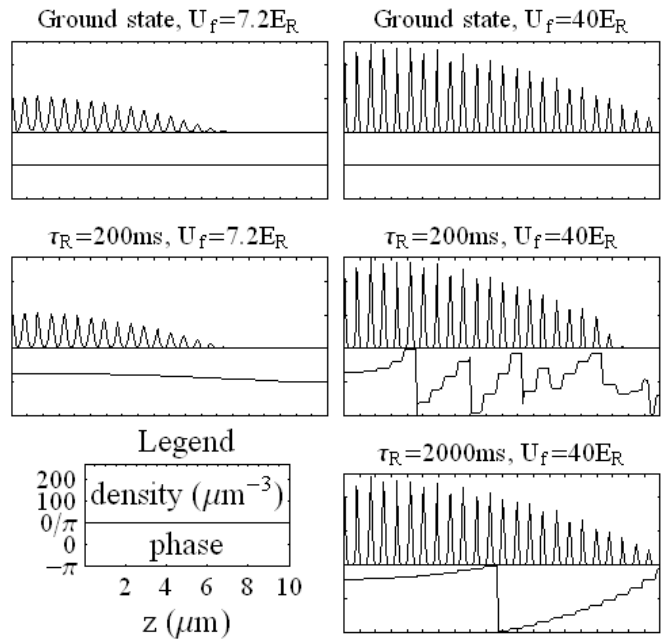}
\caption{Density profiles of ground state with the trap and lattice compared
with density profiles after turning up laser. Each plot shows the density and
phase along the longitudinal axis and $z>0$.  The first row shows the density profiles of the ground state of
the condensate in the trap and lattice for $U=7.2E_R$ and $40E_R$.  The second
row shows the density profile after creating the condensate in the trap and
turning up the laser in $\tau_R=200$~ms for the same laser strengths.  If the
laser turn-on is adiabatic, the profiles in the first and second rows should be
identical.  The profiles are nearly identical for $7.2E_R$, aside from a very
small phase sag, showing that the turn-on is adiabatic for this laser strength.
For $40E_R$ after a $200$~ms turn-on, however, the density profile is distorted
at the edges and the phase sag cycles over more than $6\pi$, showing that a
longer turn on time would be required for adiabaticity at this laser strength.
The third row shows the density profile after turning the laser up to $40E_R$
in 2~s. In this case, the density matches that of the ground state and the
phase sag is much smaller, showing that this turn-on is nearly
adiabatic.}\label{relax}
\end{figure}

We have tested the assumption of adiabaticity by simulating a $200$~ms laser
turn-on in real time. In all cases the optical lattice was turned on starting
with a numerically exact ground state condensate in the magnetic trap. The
results of these tests, summarized in Fig.~\ref{relax}, are that a $200$~ms
switching time is nearly adiabatic for $U_f=7.2E_R$, but not for $U_f=40E_R$.
Much longer times are needed for higher laser intensities not only because the
laser is turned to higher intensity, but because the higher laser fields push
more of the condensate into the outer wells; the required times for the
condensate to tunnel through higher barriers to reach these outer wells are
also longer.

As the lattice is raised, the value of the chemical potential varies from site
to site, and atoms must flow from the trap center to the periphery in order to
remain in equilibrium. For sufficiently deep lattices, the time needed to
tunnel from site to site eventually exceeds the timescale of the ramp. The
resulting non-adiabaticity for the $U_f=40E_R$ case leads to significant
deviations of both the density and phase profiles, compared with the true
ground state. First, the density envelope is truncated, reflecting the
inability of atoms to fully tunnel out to the cloud surface. Second, there
is considerable `phase sag,' illustrated in the second column of
Fig.~\ref{relax}. This phase profile, which results from the axial velocities
acquired by the atoms as they propagate (${\bf v}\propto\nabla\varphi$),
eventually becomes locked in for very deep lattices.

A simple estimate of the timescale required for adiabaticity is
that it should at least exceed the inverse of the smallest collective mode
frequency in the
presence of the lattice. In deep lattices where the atomic profile in a
given well is not much different from that of an ideal gas, the effective
axial frequency shifts as $\tilde{\omega}_z=\omega_z\sqrt{m/m^*}$, where
$m^*$ is the effective mass of the atom~\cite{Kramer2002}. For a $40E_R$ lattice, $m^*/m\approx 900$~\cite{Kramer2002}, which yields $2\pi/\tilde{\omega}_z\approx
600$~ms. It is reasonable to expect the full adiabatic ramp to require several
times this, $t_R\sim 2$~s, as is fully confirmed by the numerical results
shown in Fig.~\ref{relax}. This value is an order of magnitude larger than that
used in the experiment; since it is comparable to the lifetime of the BEC,
such a long ramp is probably not experimentally feasible. An alternative
protocol would have been to grow a coherent condensate ground state in the
presence of the magnetic trap and optical lattice.

After turning the laser up to $U_f$ in $200$~ms, the experimenters turn off the magnetic trap in $40~\mu$s and hold the BEC in the vertically oriented laser for approximately $2.5$~ms in order to produce a $\pi$ offset between adjacent wells.  The confinement of the laser is sufficient to prevent significant movement of the condensate during this time.  However, the gradient of the gravitational potential between neighboring wells of the lattice causes dramatic Schr\"{o}dinger phase evolution, since the condensate phase is given by $\theta=Vt/\hbar$.  Since there is a difference of potential energy between the wells equal to $\Delta V_g=mg\Delta z$, where $\Delta z=\lambda/2$ is the distance between the wells, after the condensate is held in the laser for a time $t_h$ there will be a phase difference between the condensates in two neighboring wells equal to $\Delta\theta=mg\lambda t_h/2\hbar$.  To achieve $\Delta\theta=\pi$, the hold time must be an odd multiple of $2\pi\hbar/mg\lambda=0.557$ms for the parameters used in this experiment.

In  our calculations, we follow this experimental procedure, using the exact hold time for $\Delta\theta=5\pi$,
$t_h=2.785$~ms, rather than the $2.5$~ms quoted in OTFYK~\cite{Kasevich2001a}.
Simulations with $t_h=2.5$~ms give an asymmetric density distribution after
expansion, in which one peak is about twice as large as the other. This
asymmetry is due to the asymmetry of the Fourier components of the wave
function when the phase difference between the wells is not exactly
$\pi$~\cite{Anderson1998a}.

\begin{figure}[t]
\centering
\includegraphics[width=\columnwidth]{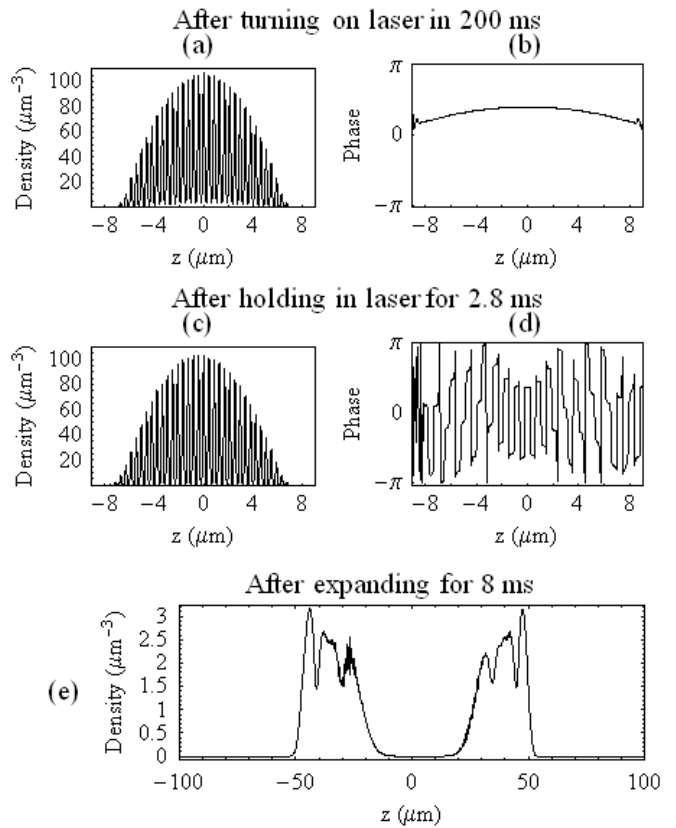}
\caption{Real time evolution for a weak laser.  (a) Density profile along
the longitudinal axis after turning laser up to $7.2E_R$ in $200$~ms.  (b) Density profile along the longitudinal axis
after turning off the magnetic trap holding the BEC in the laser (with gravity)
for $2.785$~ms.  (c) Phase profile along the longitudinal axis after turning
laser up to $7.2E_R$ in $200$~ms.  (d) Phase profile along the longitudinal
axis after turning off the magnetic trap holding the BEC in the laser (with
gravity) for $2.785$~ms. (e) Density profile along the longitudinal axis
after releasing the BEC and allowing it to expand for $8$~ms.}\label{weaklaser}
\end{figure}

\subsection{Real Time Evolution for a Weak Laser}
For a weak laser, the $200$~ms turn-on is nearly adiabatic, but it is still
important to check the results of the phase evolution and expansion of the
actual state after the laser is turned on.  The results of this calculation
are shown in Fig.~\ref{weaklaser}.  These results are as expected in that there
is approximately a $\pi$ phase difference between each neighboring well, and a
clear two-peak interference pattern after expansion.  Thus, although the
simulated dynamics of the BEC in the weak laser are not as clean as the
predictions of Fig.~\ref{expansions}, the simulations confirm that the basic
ideas of the discussion of Section III are correct in this case, and in
agreement with the OTFYK observations.

\begin{figure}[t]
\centering
\includegraphics[width=\columnwidth]{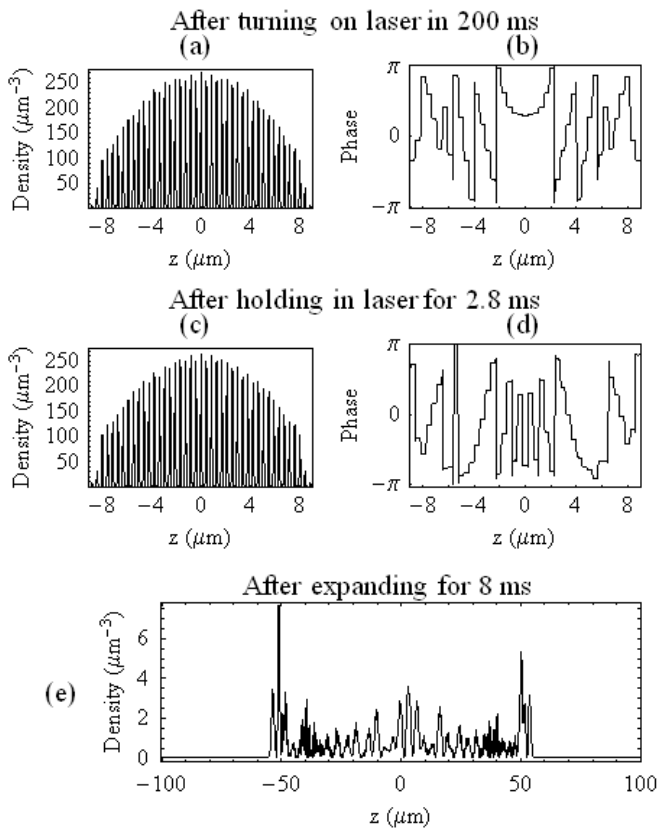}
\caption{Real time evolution for a strong laser.  (a) Density profile
along the longitudinal axis after turning laser up to $40E_R$ in $200$~ms.
(b) Density profile along the longitudinal axis after turning off the
magnetic trap holding the BEC in the laser (with gravity) for $2.785$~ms.
(c) Phase profile along the longitudinal axis after
turning laser up to $40E_R$ in $200$~ms.  (d) Phase profile along the
longitudinal axis after turning off the magnetic trap holding the BEC in the
laser (with gravity) for $2.785$~ms. (e) Density profile along the
longitudinal axis after releasing the BEC and allowing it to expand for
$8$~ms.}\label{stronglaser}
\end{figure}

\subsection{Real Time Evolution for a Strong Laser}
Fig.~\ref{stronglaser} illustrates the effect of ramping the lattice up to
40$E_R$ in 200~ms. The resulting phase sag shown in the second column of
Fig.~\ref{relax} is repeated in Fig.~\ref{stronglaser}(c). After holding the
atoms in the laser for $2.785$~ms, as shown in Fig.~\ref{stronglaser}(d), the
phase is so distorted that it begins to resemble the random pattern used
in Fig.~\ref{expansions}.  In this context it is not surprising that the
density profile after expansion, depicted in Fig.~\ref{stronglaser}(e), is
completely washed out. The loss of interference is driven entirely by
non-adiabatic effect captured within a mean-field model. It is important to
underline that these non-adiabatic effects are intrinsic to the ramp time,
and are not affected by how the ramp is applied. We have performed simulations
using a ramp with a smooth onset (based on a sine function) rather than the
linear ramp discussed above, but the observed loss of interference was
unchanged. We are thus able to qualitatively reproduce the loss of interference without invoking number squeezing.

\subsection{Absorption Probabilities}\label{secabsorption}
To compare the results of our simulations directly with the experimental data,
we calculated absorption probabilities, which we then smoothed to account for
finite experimental resolution.  For each simulation, we integrate the 3D
density profile after $8$~ms of expansion, $N|\psi(x,y,z)|^2$, over one of the transverse
axes (the direction of the imaging beam), to get the integrated density:
\begin{equation}\label{ntilde}
\tilde{n}(x,z)=\int dy |\psi(x,y,z)|^2
\end{equation}
We then calculate the absorption probability from the following equation
\cite{Ketterle1999a}:
\begin{equation}\label{A}
A(x,z)=1-\exp\left(-\sigma_0 \tilde{n}(x,z)\right)
\end{equation}
where the absorption cross section $\sigma_0$ is given by:
\begin{equation}\label{crosssection}
\sigma_0=6\pi\lambdabar^2
\end{equation}
with $\lambdabar=780$nm$/2\pi$.
We then smooth the data by taking a Gaussian convolution:
\begin{equation}\label{smoothcrosssection}
\tilde{A}(x,z)=\int dx' dz' A(x',z')e^{-((x-x')^2+(z-z')^2)/w^2}
\end{equation}
where the $1/e$ width $w=17\mu$m is chosen so that at very low laser strength
($U_f=5-8E_R$) the ratio of the width of the interference peaks to their
separation is $\zeta=0.22$. This value of $\zeta$ corresponds to the best
experimental contrast reported in Ref.~\cite{Orzel2001a} (the parameter
$\zeta$ will be discussed further in Sec.~\ref{seczeta}).  The same method was
used by OTFYK to analyze their experimental results, but they found that they
needed a $1/e$ width of $25\mu$m, rather than $17\mu$m, to match their
data~\cite{Tuchman2004b}.

\begin{figure}[t]
\centering
\includegraphics[width=\columnwidth]{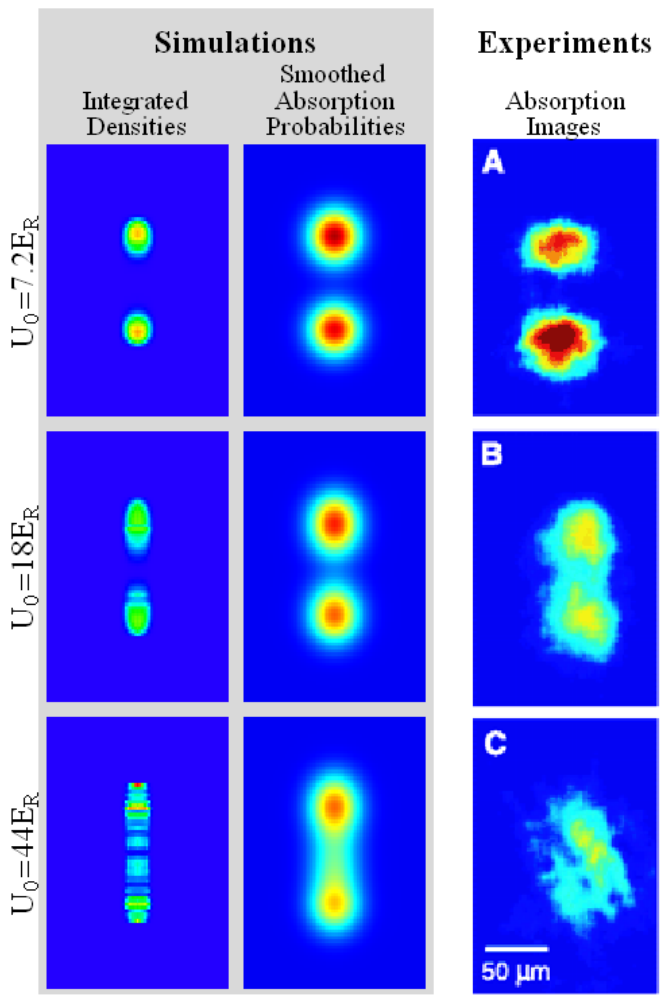}
\caption{Comparison between experimental results and numerical simulations.
Interference patterns after turning laser intensity up to $U_f$ in $200$~ms,
turning off trap and holding BEC in laser for $2.785$~ms, then releasing it and
allowing it to expand for $8$~ms. (a) $U_f=7.2E_R$ (b) $U_f=18E_R$ (c)
$U_f=44E_R$}\label{kasevichcomparison}
\end{figure}

Fig.~\ref{kasevichcomparison} shows how the results of time-dependent numerical
simulations compare with the experimental results. The three columns depict
the raw numerical results for integrated density, the smoothed absorption
probabilities, and the experimental results (copied with permission from
the authors), respectively.

The numerical simulations can reproduce qualitatively the observed
interference patterns for both weak and strong lasers.  The flat phase
profile after loading the atoms into a $7.2E_R$ lattice guarantees a clean
interference pattern~\cite{Pedri2001}. For a
strong laser, the underlying mechanism for the loss of interference in our
simulations (phase distortion due to non-adiabatic mean field effects) is
entirely different from the mechanism proposed by the experimenters (number
squeezing).  A more quantitative analysis, given in Section VIII, reveals that
the correspondence between simulation and experiment is not perfect, so it may be
that number squeezing is in fact occurring in the experiment.  However, the
results of these simulations show that loss of interference is not in itself
sufficient evidence of the presence of number squeezing.

\begin{figure}[t]
\centering
\includegraphics[width=\columnwidth]{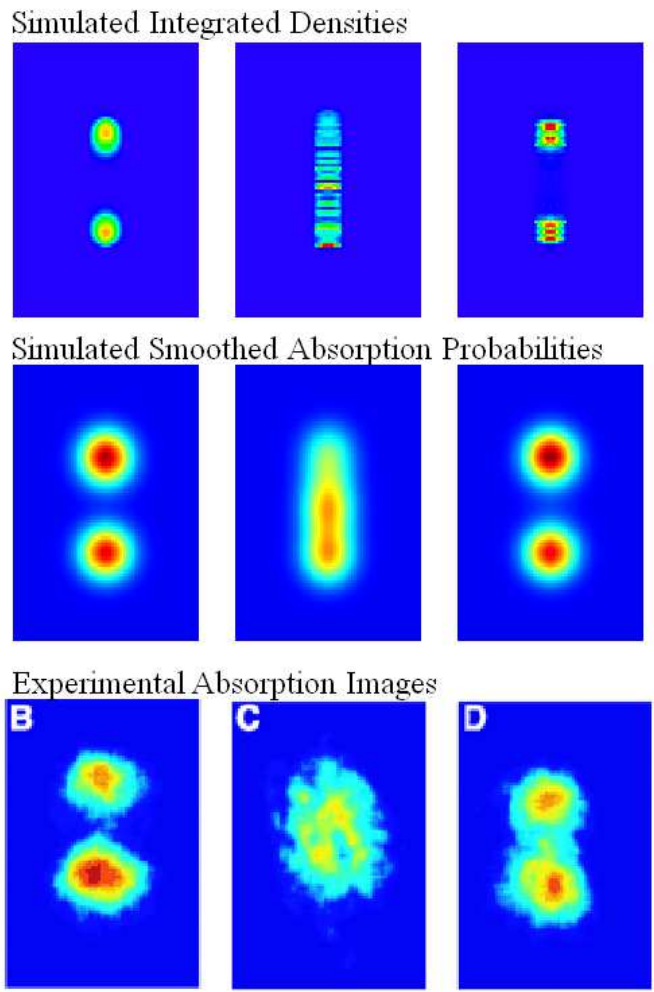}
\caption{Restoring interference: Comparison between numerical and experimental
results. (b) Interference pattern after releasing BEC from $7.2E_R$ laser. (c)
Interference pattern after turning laser up to $40E_R$ and releasing BEC. (d)
Interference pattern after turning laser up to $40E_R$, then turning laser down
to $10E_R$, and then releasing BEC.}\label{restorecomparison}
\end{figure}

\section{Restoring Interference}
\begin{figure}[t]
\centering
\includegraphics[width=\columnwidth]{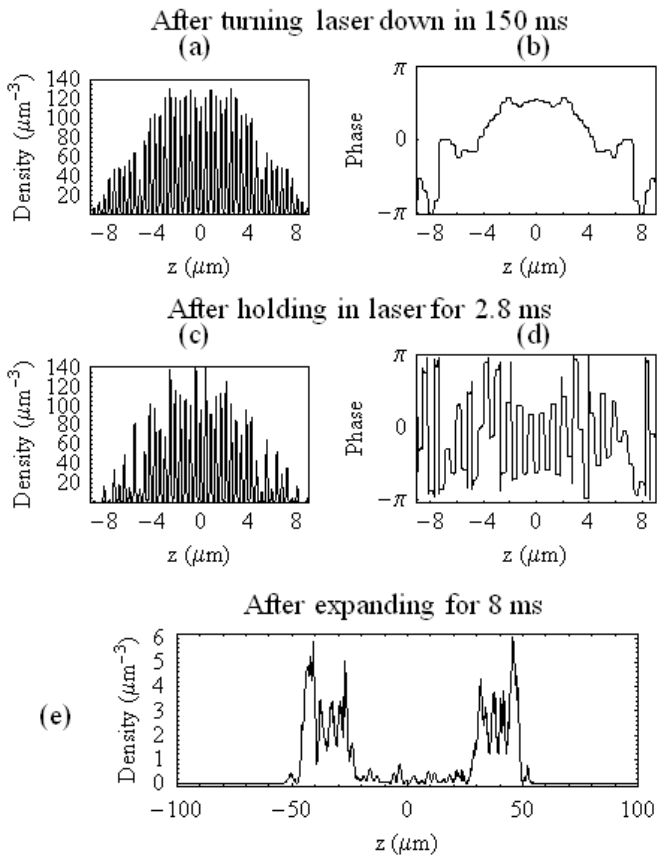}
\caption{Real time evolution after turning laser up and then down.  (a) Density profile along the longitudinal axis after turning laser up to $40E_R$
in $200$~ms and then down to $10E_R$ in $150$~ms.  (b) Density profile
along the longitudinal axis after turning off the magnetic trap holding the
BEC in the laser (with gravity) for $2.785$~ms.  (c) Phase profile along
the longitudinal axis after turning
laser up to $40E_R$ in $200$~ms and then down to $10E_R$ in $150$~ms.  (d) Phase profile along the longitudinal axis after turning off the magnetic trap
holding the BEC in the laser (with gravity) for $2.785$~ms. (e) Density
profile along the longitudinal axis after releasing the BEC and allowing it to
expand for $8$~ms.}\label{downdynamics}
\end{figure}

In the next phase of the experiment, instead of releasing the BEC  after
turning the laser up to $40E_R$, OTFYK first turn the laser down to $10E_R$
in $150$~ms and then release the condensate after the short hold time.  In the
experiments the interference pattern, and by implication the coherence of the
BEC, is restored by the adiabatic ramp-down. The results were originally
explained in terms of the time-dependent two-mode model as the loss and return
of coherence of the BEC.  However, Fig.~\ref{restorecomparison} illustrates
that they can also be reproduced with numerical simulations of the GPE, and
can therefore be explained in terms of mean-field effects. Indeed, the
original interpretation of these experiments in terms of number squeezing has
since been revised~\cite{Tuchman2004a}. The dynamics are
illustrated in Fig.~\ref{downdynamics}, which shows the density and phase
profiles after turning down the laser, before and after releasing the BEC.
This figure shows that after turning the laser up and then down, the phase sag
unwinds, so that at the end of the process, the site-to-site phase is
relatively constant.  There are variations in phase, but their size is on the
order of $\pi$, rather than $7\pi$.  There is considerably more noise in the
system after this process than there would have been if the laser were simply
turned up to $10E_R$ without first going through the high barrier state, but
this noise does not obscure the basic pattern.

\section{Random Phase Imprinting}
A further step, which was not done in the experiment, but which in principle
could be done, is to imprint a random phase shift on each well when the laser
strength is $40E_R$ and then turn the barrier back down.  The two-mode model
predicts that applying a random phase shift will destroy the ability of a
mean-field state to heal back to a stationary (flat-phase) BEC state,
but will have no effect on a strongly number-squeezed state.  Since we
have shown that the loss of interference is not sufficient to demonstrate
number squeezing, it appears that random phase shifts could be used as an
alternative test.  If this is to be an effective test, applying a random phase
shift should completely destroy the ability of the resulting BEC to
produce a clean interference pattern.

\begin{figure}[t]
\centering
\includegraphics[width=\columnwidth]{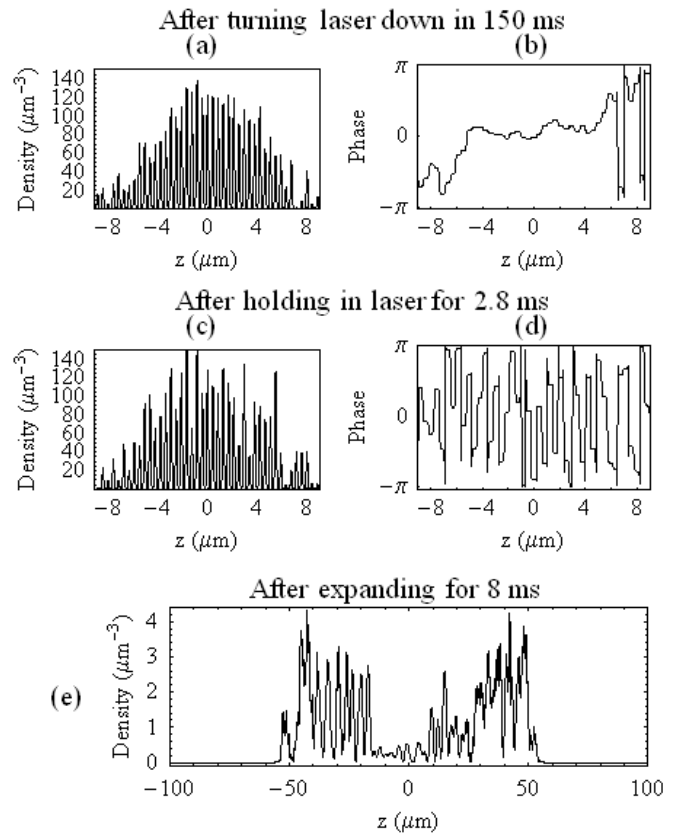}
\caption[Real time evolution after turning laser up, applying a random phase
shift to each well, and then turning laser down]{Real time evolution after
turning laser up, applying a random phase shift to each well, and then turning
laser down.  (a) Density profile along the longitudinal axis after laser is
turned down.  (b) Density profile along the longitudinal axis after turning
off the magnetic trap holding the BEC in the laser (with gravity) for
$2.785$~ms.  (c) Phase profile along the longitudinal axis after turning
laser up to $40E_R$ in $200$~ms and then down to $10E_R$ in $150$~ms.  (d) Phase profile along the longitudinal axis after turning off the magnetic trap
holding the BEC in the laser (with gravity) for $2.785$~ms. (e) Density
profile along the longitudinal axis after releasing the BEC and allowing it to
expand for $8$~ms.}\label{randdynamics}
\end{figure}

\begin{figure}[t]
\centering
\includegraphics[width=\columnwidth]{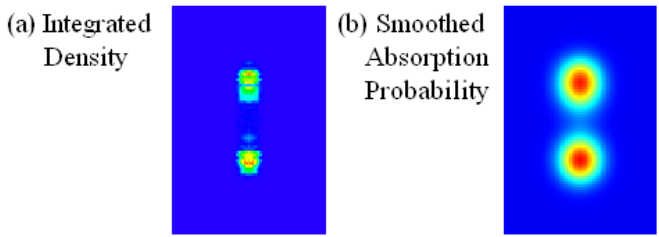}
\caption{Integrated density and absorption probability of expanded BEC after
turning laser up, applying a random phase shift to each well, and then turning
laser down}\label{randsmooth}
\end{figure}

We tested this idea numerically by running simulations similar to those
described in the previous section, where we turned the laser up to $40E_R$ and
then down to $10E_R$, but in this case we applied a random phase shift to each well before
turning the laser down.  The surprising result, illustrated in
Fig.~\ref{randdynamics}, is that the non-linear mean field dynamics
`self-heals' a truly random phase pattern just as it unwinds a mean-field
induced phase sag:  the interference pattern is nearly as dramatic as in the
simulations in the previous section where there was no phase shift.  In
Fig.~\ref{randdynamics}, the two-peak interference pattern is not very clean,
but
Fig.~\ref{randsmooth} shows that after smoothing, the absorption probability is
not significantly different from that acquired after turning the laser up and down
without a phase shift.  Thus, even in the case of random phase shifts, mean
field effects are able to mimic the predictions of the two-mode model and it
appears to be difficult, under the OTFYK conditions, to distinguish
experimentally between mean field effects and number squeezing.

\section{Phase Oscillations}\label{secphaseoscillations}
\begin{figure}[t]
\centering
\includegraphics[width=\columnwidth]{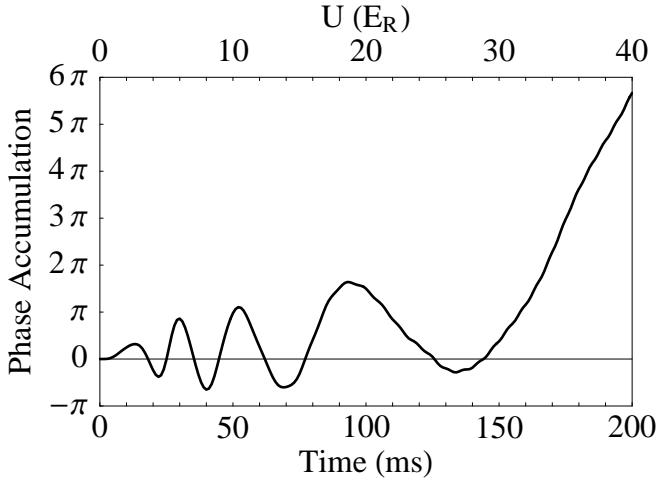}
\caption{Phase accumulation, defined as phase difference along the longitudinal axis between $z=6~\mu$m and $z=0~\mu$m (after unwrapping), plotted as a function of time, as the laser is turned up to $U_f=40E_R$ in $200$~ms.}
\label{phaseoscillations}
\end{figure}

An unexpected effect observed in the numerical simulations during the turn-on
of the laser is a slow oscillation of the phase.  The phase sag does not
continue to grow, as has been seen in previous simulations using effective 1D
models~\cite{Sklarz2002,Band2002b}, and which one might expect if the phase
dynamics were due only to the mean field potential differences between the
wells.  Instead, as shown in Fig.~\ref{phaseoscillations}, the phase
oscillates, with the sag growing, then shrinking, then reversing.

\begin{figure}[t]
\centering
\includegraphics[width=\columnwidth]{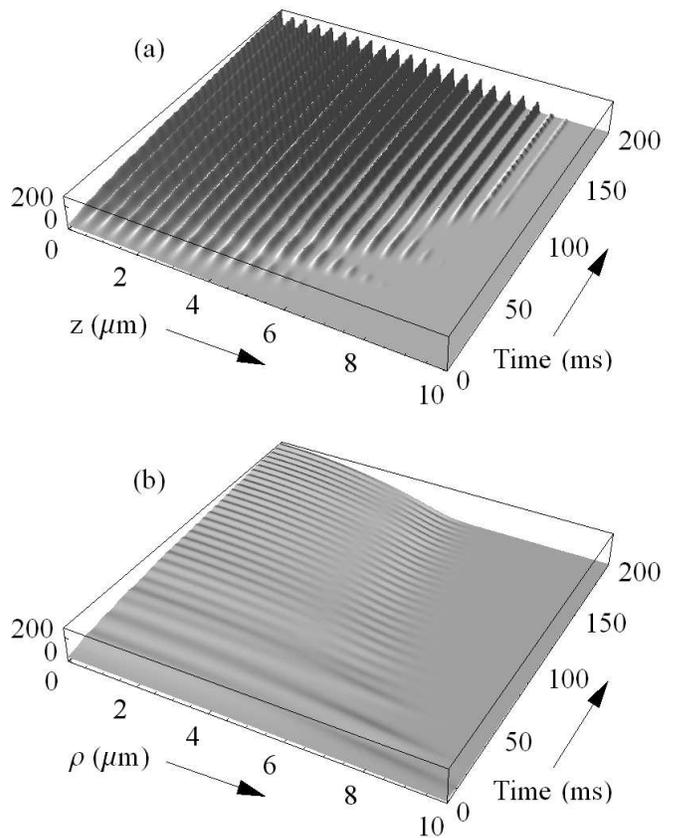}
\caption{Density oscillations. (a) Density (in units of $\mu$m$^{-3}$) along
half the longitudinal axis as a function of time as laser is turned up to
$40E_R$.  (b) Density (in units of $\mu$m$^{-3}$) along half the transverse
axis as a function of time as laser is turned up to $40E_R$ in $200$~ms.  On the whole the
BEC is getting wider in the longitudinal direction and narrower in the
transverse direction, but the width oscillates in both directions along the
way.  The oscillations are very large for about the first 100~ms, and much
smaller after that.}\label{densityoscillations}
\end{figure}

These phase oscillations are the result of two non-adiabatic effects. First,
if the lattice is ramped up too quickly, the local chemical potentials
$\mu_{\rm loc}$ in each well, defined by the expectation of the GPE operator,
will not be identical. At the end of the ramp, the phase in each well will
vary in time as $\varphi_{\rm loc}\sim\mu_{\rm loc}t/\hbar$, causing the
overall phase profile to vary in time. The second, more important, source of
the oscillations is the induction of transverse and longitudinal collective
modes of the condensate.  As the laser is turned up, the transverse confinement
increases (see Fig.~\ref{groundstates}), and the BEC is pulled inward in the
transverse direction and pushed outward in the longitudinal direction.  This
flow of superfluid gives rise to density oscillations along the longitudinal
and radial axes, as illustrated in Fig.~\ref{densityoscillations}. As the
lattice deepens, the frequency of the axial oscillations decreases because of
the increasing effective mass~\cite{Kramer2002} and the time between classical
turning points (where the phase is most flat) lengthens.

For a sufficiently deep lattice, the
axial dipole mode period exceeds experimental timescales and the
accumulated phases become effectively locked in. If the radial modes in each
well all have the same frequency and are in phase, then the axial phase profile
would be unaffected by their presence. In fact, for non-adiabatic ramps each
disconnected well has a slightly different radial confinement frequency, so
one would expect the longitudinal phase profile to change with time even
for deep lattices. As discussed in the next section, these phase oscillations
give rise to `collapses and revivals' of the interference pattern which were
not observed in the OTFYK experiments.

\section{Zeta and Visibility}\label{seczeta}

\begin{figure}[t]
\centering
\includegraphics[width=\columnwidth]{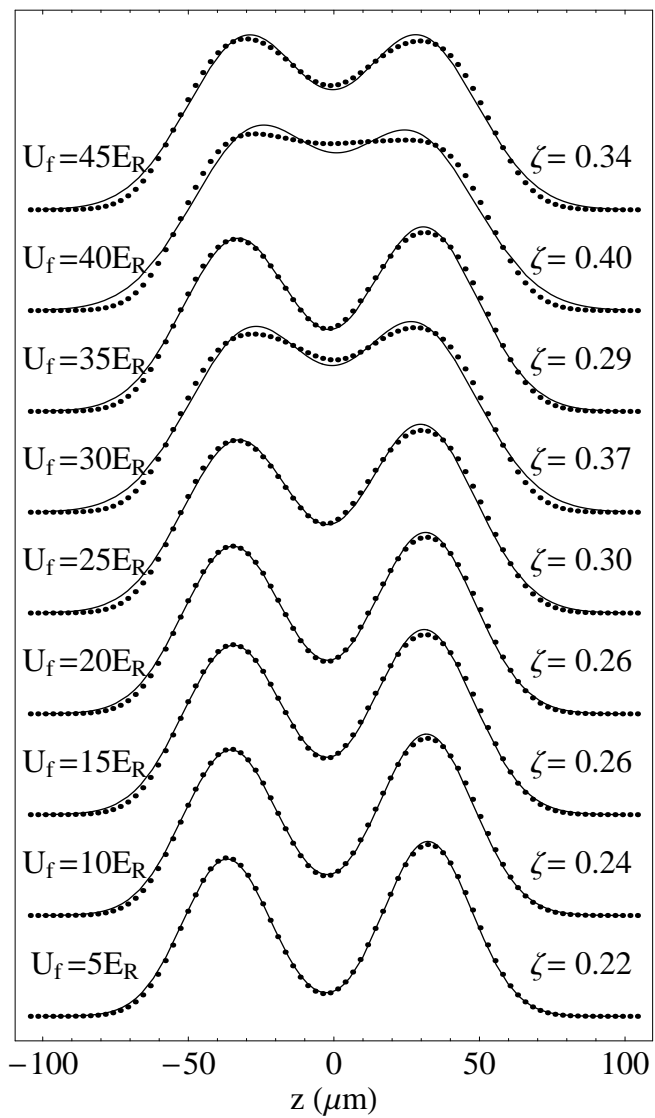}
\caption{Smoothed absorption cross sections (dots) and double Gaussian fits
(solid lines) for a range of laser strengths.  The fits are not perfect,
especially at high laser strengths, but the differences between the fits and the
data are much smaller than the differences between the data at different laser
strengths.  The fit shown for $40E_R$ is the worst fit in the data
set.}\label{fits}
\end{figure}

For more quantitative comparisons with experiment, it is useful to calculate
the quantities used by OTFYK to compare their number squeezing model, their
effective one dimensional GP model, and their data, namely $\zeta$ and
visibility \cite{Tuchman2004a,Tuchman2005}.  $\zeta$ is defined as the ratio
of the width
of a single peak to the distance between the peaks and $\zeta_0=0.22$ is the
value of $\zeta$ for $U_f=6E_R$.  To determine $\zeta$, we fit the cross
section of the smoothed absorption probability through the longitudinal axis
$z$, $\tilde{A}(0,z)$, to a double Gaussian:
\begin{equation}\label{doublegaussian}
Be^{-(z-z_1)^2/2\sigma^2}+Ce^{-(z-z_2)^2/2\sigma^2}
\end{equation}
and then $\zeta$ is the ratio of the width of the Gaussians to the distance
between their centers:
\begin{equation}\label{zetaEqn}
\zeta=\sigma/(z_1-z_2)
\end{equation}
and $\zeta_0$ is the value of $\zeta$ for $U_f=6E_R$.  Fig. \ref{fits} shows a
sample of smoothed absorption cross sections and double Gaussian fits for a
range of laser strengths.  Visibility is defined as the difference between the
average of the maxima of the two peaks and the minimum between the peaks,
divided by their sum.

In a further analysis of the OTFYK experiment, Tuchman~\cite{Tuchman2004a}
plots $\zeta$ and visibility as a function of $Ng\beta/\gamma$, where $N$ is
the is number of particles in two wells, $g$ is defined as in Eq.~(\ref{gpe}),
and $\beta$ and $\gamma$ are defined by the following integrals over localized
wave functions:
\begin{eqnarray}
\label{beta}\beta\equiv\int d^3r\psi_1^4(\mathbf{r})\\
\label{gamma}\gamma\equiv\int d^3r\psi_1(\mathbf{r})\{-
\frac{\hbar^2}{2m}\nabla^2+V^{ext}(\mathbf{r})\}\psi_2(\mathbf{r})
\end{eqnarray}

Unlike $\zeta$ and visibility, which are determined by fits to experimental data, the parameter $Ng\beta/\gamma$ is derived from a two-mode model in which there are two fixed wave functions, $\psi_1$ and $\psi_2$, in each of two potential wells in the optical lattice.  This parameter is essentially a measure of laser strength and density, and it is approximately exponential in laser strength.  Our own calculations show that this factor varies somewhat depending on the details of the theoretical model used to determine it.  Therefore, we prefer to plot $\zeta$ and visibility versus laser strength $U_f$, an experimentally determined parameter, rather than versus $Ng\beta/\gamma$, which is based on particular theoretical model.  According to the numbers given in Ref.~\cite{Orzel2001a}, a range of laser strengths from $6E_R$ to $50E_R$ corresponds approximately to a range of $Ng\beta/\gamma$ from $10^{1/2}$ to $10^5$.

\begin{figure}[t]
\centering
\includegraphics[width=\columnwidth]{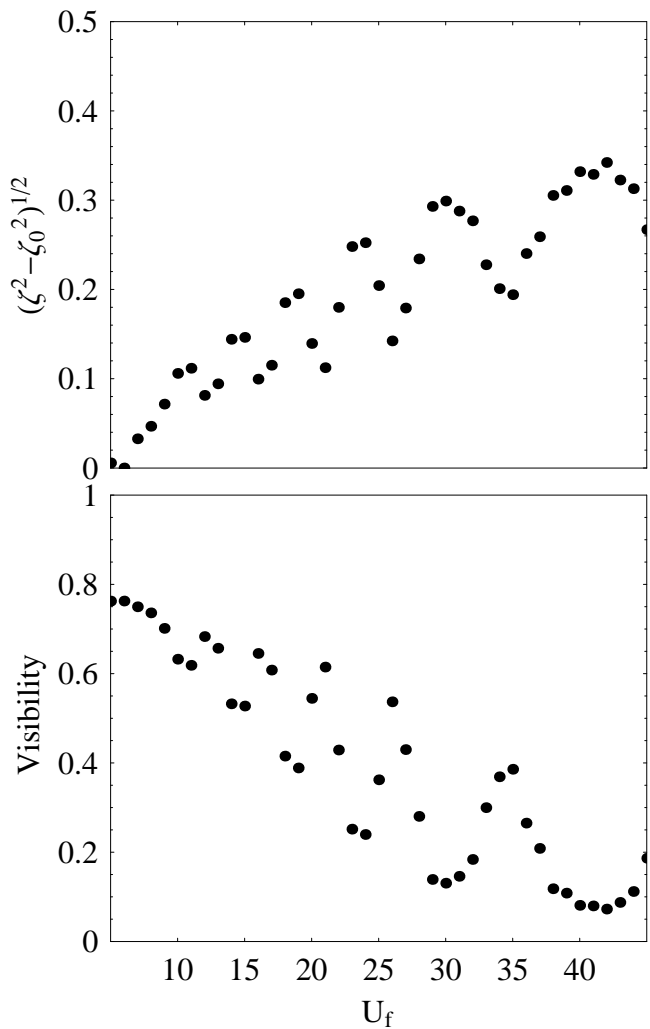}
\caption{$(\zeta^2-{\zeta_0}^2)^{1/2}$ and visibility versus $Ng\beta/\gamma$
and laser strength.}\label{zetavis}
\end{figure}

Fig.~\ref{zetavis} shows $\zeta$ and visibility for our simulations as a
function of $U_f$ (laser strength upon release).  Each point
corresponds to a simulation in which the laser strength is raised to $U_f$ in
200~ms, the magnetic trap is turned off in $40~\mu$s, the BEC is held in the
laser for 2.785~ms, the laser is turned off and the BEC is allowed to expand
for 8~ms. For each point, $\zeta$ and visibility are calculated from a cross
section of the smoothed absorption probability discussed in
Sec.~\ref{secabsorption}.

Perhaps the most striking aspect of the numerical results is that the
values of both $\zeta$ and visibility do not merely increase or decrease as a
function of $U_f$, but oscillate.  These oscillations are due to the excitation
of collective modes as discussed in the previous section, and do not appear in
the experimental data or in the models employed in
Refs.~\cite{Orzel2001a,Tuchman2005}.  It is possible that the oscillations
present in the GPE simulations would be strongly damped in actual experiments,
with the high radial energies being transferred to high-lying axial modes. When
atoms are loaded into blue-detuned lattices, where the transverse
confinement due to the external magnetic trap is weak compared to that of
red-detuned lattices, considerable radial heating has been
observed~\cite{Morsch2003} leading to loss of interference
contrast~\cite{Hadzibabic2004}.

The observation of phase oscillations in numerical simulations leads to
predictions that would be interesting to test experimentally. If the BEC were
released when there was a peak in the amplitude of the phase oscillations, the
interference pattern could be lost for a relatively low barrier height.  If
one waited a little longer to release the BEC until the phase flattened out
again, the interference pattern would return for a few oscillations until
disappearing due to damping. Note that these `collapses and revivals' are
completely driven by mean-field effects, and are unrelated to similar phenomena
found in 3D lattices~\cite{Greiner2002b}. Such effects were not reported in
the OTFYK work, but contrast oscillations attributed to quantum fluctuations
were recently observed by the same group after non-adiabatically ramping a deep
lattice to a relatively shallow depth of 16.6$E_R$~\cite{Tuchman2005}.

While we cannot compare the results shown in Fig.~\ref{zetavis}
directly to the experimental results because we are plotting our results
as a function of $U_f$ rather than $Ng\beta/\gamma$, we can make a
qualitative comparison based on the given ranges of $U_f$ and
$Ng\beta/\gamma$.  Our results are qualitatively similar to those found
in experiments~\cite{Tuchman2004a}, in that the fringe contrast
decreases approximately linearly with lattice depth. With our best
estimate of the correspondence between $Ng\beta/\gamma$ and $U_f$, the
average values of $\zeta$ found in the simulations are consistently
smaller (on the order of 30\%) than the experimental data. Besides this
uncertainty, this systematic difference is likely due to at least two
factors.  First, the mean-field calculations cannot incorporate the
physics of number squeezing that is likely to be a contributing factor
in real systems. Second, the damping of the induced collective modes
discussed above should diminish the effective scatter of the data points
(of order $25\%$ of the mean values shown in Fig.~\ref{zetavis}),
heating the system and reducing the overall interference visibility.

\section{Conclusions}
We have shown that mean-field effects, as modeled by the three-dimensional
time-dependent GPE, can explain both the loss and return of interference for a
BEC in a one-dimensional optical lattice. The simulations yield behavior that
is qualitatively similar to that observed of the OTFYK
experiments~\cite{Orzel2001a} without the need to invoke physics beyond
mean-field theory, such as number
squeezing or condensate fragmentation. The central result of
the present work is that ensuring adiabaticity during the loading of a BEC
into a deep optical lattice is experimentally difficult to achieve, and
that neglecting the rather large effects of mean-field excitations will
lead to an incomplete description of future experiments.

The results of GPE simulations differ from the experimental data in a few
small respects, however. The loss of contrast in the interference patterns,
after dropping a deep optical lattice and allowing the cloud to expand, is
not quite as pronounced as in the experiments. One explanation for the
small difference is likely to be the presence of number squeezing, which is
not captured by the GPE. Another possibility is that the radial excitations
that are induced by the lattice ramp will rapidly damp into high-lying
axial excitations in real experiments (the GPE has no damping mechanism),
leading to heating and its attendant
loss of contrast. These collective modes lead to phase oscillations that
in turn yield periodic variations in the fringe contrast; though these
`collapses and revivals' were not seen in the original OTFYK experiments,
similar oscillations attributed to quantum fluctuations have been observed
more recently by the members of the same group~\cite{Tuchman2005}. Because the
creation and destruction of interference patterns are widely used
techniques to measure the presence of coherence and its loss, it is important
to understand all the mechanisms that can affect this important experimental
measure.

\section{Acknowledgments}
The authors would like to thank Mark Kasevich and Ari Tuchman for many useful
discussions, the sharing of unpublished data and their own computational
results, and for their most generous hospitality in hosting S.~B.~McKagan's
visit to Stanford for an extended discussion of the present work. The
enthusiastic support of Charles Clark at NIST is also gratefully acknowledged.
This work was supported by NSF grant PHY-0140091 (WPR \& SBM), a Dissertation
Fellowship from the American Association of University Women (SBM), and by
the Natural Sciences and Engineering Research Council of Canada (DLF).

\appendix

\section{Numerical Methods}\label{numericalmethods}
The numerical simulation of the full three dimensional time-dependent
Gross-Pitaevskii Equation (GPE) is performed in Cartesian coordinates using a C
code adapted from earlier work in the Reinhardt group.  The time integration
uses the variable step fourth-fifth order Runge-Kutta integrator {\tt odeint}
from Numerical Recipes in C~\cite{Press1992a}.  The spatial integration uses a
pseudospectral method~\cite{Fornberg1996a} with fast Fourier transforms from
the fftw library~\cite{fftw}.  The basic idea behind the pseudospectral method
is that the wave function is expanded in terms of coordinate discretized
trigonometric functions, reducing a partial differential equation into a set
of coupled ordinary differential equations via fast Fourier transforms to
coefficient basis.  We have adapted the code using Type II Fourier transforms,
which are appropriate to
the boundary conditions of half of a symmetric box, to take advantage of the
symmetry of the problem and reduce the analysis to one quadrant of the system.
This adaptation increases the speed of the calculations by a factor of 8.  In
theory, since the system of interest is cylindrically symmetric, further
improvement in speed could be achieved by using cylindrical coordinates and
thus effectively reducing the problem two dimensions.  In practice, the speed
of the
fast Fourier transforms appropriate to Cartesian coordinates eliminate the
disadvantage of working in 3D and the 3D Cartesian code runs significantly
faster than a 2D cylindrical coordinate code using the discrete variable
representation method rather than the pseudospectral method.

For the results shown in this paper, we use 24.86 grid points per micron in
the longitudinal direction (10.44 grid points per site) and 1.09 grid points
per micron in the transverse directions.  During the expansion of the BEC, we
reduce the number of grid points per unit length in the longitudinal direction
by a factor of 2 during the first 4~ms, and then by an additional factor of 2
during the next 4~ms. We have
done sample runs with up to twice as many longitudinal grid points and four
times as many grid points in each transverse direction and checked that this
does not change the results.  Many more grid points are needed in the
longitudinal direction than in the transverse direction because of there is
much more variation due to the laser, which has a wavelength of $0.84\mu$m.

The simulations were done on a computer with four 3~GHz Xeon processors and
2~Gb of memory running Red Hat Linux.  On this machine, using the Intel C
compiler, the computations of the BEC expansion take about 4 hours using the
minimum necessary number of grid points and the 200 ms laser turn-on
computations take about 40-60 hours.  (The code runs about 1.5 times faster
when compiled with the Intel C compiler than when compiled with gcc.)

\bibliography{paper}

\begin{thebibliography}{67}
\bibitem[{com()}]{comment}
\bibinfo{note}{The experiment also involved some runs in which the laser was
  turned up quickly, but these will not be discussed.}
\end{thebibliography}

\end{document}